\documentclass{article}
\usepackage{spconf,amsmath,graphicx}

\usepackage{soul,color}
\usepackage{enumitem}
\usepackage{url}
\usepackage{subcaption}
\usepackage{booktabs} 

\newcommand{\compresslist}{%
    \setlength{\topsep}{\parskip}%
    \setlength{\itemsep}{1pt}%
    \setlength{\parskip}{0pt}%
    \setlength{\parsep}{0pt}%
}

\setlist[itemize]{leftmargin=*}

\def\musicnn{MusiCNN}

\title{TensorFlow Audio Models in Essentia}

\name{Pablo Alonso-Jim\' enez, Dmitry Bogdanov, Jordi Pons, Xavier Serra}
\address{Music Technology Group, Universitat Pompeu Fabra}

\begin{document}
\ninept
\maketitle
\begin{abstract}


Essentia is a reference open-source C++/Python library for audio and music analysis. In this work, we present a set of algorithms
that employ TensorFlow in Essentia, allow  
predictions with 
pre-trained
deep learning 
models, and are designed to offer flexibility of use, 
easy extensibility, and real-time inference. 
To show the potential of this new interface with TensorFlow, we provide a number of pre-trained state-of-the-art  
music tagging and classification CNN models. 
We run an extensive evaluation 
of 
the developed models. 
In particular, 
we assess the generalization capabilities 
in a cross-collection evaluation utilizing both 
external tag datasets as well as manual annotations tailored to the taxonomies of our models. 

\end{abstract}
\section{Introduction}
\label{sec:intro}

Audio signal processing and music information retrieval (MIR) have significantly evolved with recent advances in deep learning. As a result, many existing software tools for audio analysis are 
lacking the functionalities required by the latest state of the art and/or cannot be connected straightforwardly with external software for deep learning, especially in the case of industrial deployment. For example, a typical pipeline for 
an audio 
tagging system may include computation and pre-processing of audio features (e.g., spectrograms) using audio analysis libraries 
(e.g., Essentia, Librosa, openSMILE or Madmom~\cite{bogdanov2013essentia,mcfee_librosa:_2015,eyben2010opensmile,madmom}),  
followed by deep learning frameworks for model inference 
relying on those features (e.g., TensorFlow or PyTorch~\cite{abadi2016tensorflow,paszke2017automatic}). While 
all software in the pipeline 
may provide APIs in different languages, such as C++ and Python, and technically can be interconnected, there is a lack of efficient cross-platform software libraries incorporating all the steps in a unified pipeline to make its deployment and usage in 
applications as easy and efficient as possible. 
Some efforts have been devoted by TensorFlow (with \textit{tf.signal}) and PyTorch (with \textit{torchaudio}) to incorporate 
audio signal processing layers that can run on GPUs. Still, many deep learning practitioners rely on music/audio-specific pre-processing libraries many of which are not optimized for efficiency.


Essentia\footnote{\url{https://essentia.upf.edu}} is an open-source library for audio and music analysis released under the AGPLv3 license and well known for its capability to serve as a basis for large-scale industrial applications as well as a rapid prototyping framework~\cite{bogdanov2013essentia}. Some of its key features are: 
\begin{itemize}\compresslist
\item It is implemented in C++, with a great focus on efficiency, which makes it the fastest open-source library with the largest amount of features for audio analysis~\cite{moffat_evaluation_2015}. 
\item It supports a declarative approach to the implementation of signal processing pipelines with the {``streaming mode"} connecting algorithms for each computation step via ring buffers. This allows the user to streamline audio analysis processing input files or audio streams by chunks (in particular in real-time) and also limits memory usage, which can be crucial for many applications.
\item It has a Python interface. Programming in an interpreted language while all the data flow is ultimately controlled by optimized C++ code provides a 
balance between functionality and flexibility.
\item It supports various platforms including Linux, Windows, MacOS, Android, iOS, and can be also cross-compiled to JavaScript. 
\end{itemize}


Given its focus on efficiency, flexibility of use, modularity and easy extensibility, we consider Essentia an attractive infrastructure to build 
efficient and modular deep learning pipelines for audio. 
A similar effort in the past led Essentia to integrate a collection of Support Vector Machine (SVM) classifiers based on engineered features and trained on in-house music collections (datasets) available at Music Technology Group (MTG).\footnote{\url{https://acousticbrainz.org/datasets/accuracy}}
These classifiers are publicly available and have been used extensively for research~\cite{wack_music_2009, wack_music_2010, laurier2011, bogdanov_unifying_2011,bogdanov_semantic_2013,fricke_computer-based_2018} and in AcousticBrainz, an open database of music audio features~\cite{porter_acousticbrainz:_2015} with over 13.5 million analyzed tracks. 
These models achieved very competitive results according to a standard cross-fold validation, but when some of the classifiers were assessed in the context of external data they showed very poor performance revealing low generalization capabilities~\cite{bogdanov2016cross}. 
In addition, recent studies suggest that new approaches based on deep learning are able to outperform SVMs in audio tagging tasks~\cite{pons2017end,lee2018samplecnn,hershey2016cnn,pons2019musicnn}. For these reasons, our goals are to 
implement a new set of algorithms in Essentia and develop new classifier models based on deep learning and capable of better generalization, which can be used for both research and industrial applications.




Unfortunately, deep learning models require large amounts of training data to perform well~\cite{pons2017end,hershey2016cnn,pons2019training} and, in most scenarios, it is unreasonable to assume that large training databases are available. Considering that many Essentia use cases might be limited by the size of the datasets at hand, we limit our experiments to such cases and train models on small in-house datasets previously used for training SVM classifiers. Several studies have revealed the potential of transfer learning techniques for small training data in the context of audio auto-tagging~\cite{pons2019training,choi2017transfer}. 
For this reason, 
we investigate the generalization capabilities of this approach on our datasets. 

In short, 
transfer learning takes advantage of the knowledge acquired on an external (source) task, where more training data is available, to improve performance on the target task where data is scarce.
Generally, this is done by fine-tuning the pre-trained model~\cite{pons2019training,choi2017transfer} or by using it as a (fixed) feature extractor~\cite{lee2018samplecnn,pons2019musicnn}. In our work, 
we opt for the latter and compare 
such transfer learning models with \textit{(i)} deep learning models trained from scratch and \textit{(ii)} the SVM-classifiers based on engineered audio features. 

The rest of the paper is structured as follows: we first introduce the algorithms we have developed to integrate TensorFlow in Essentia and present a number of state-of-the-art CNN models available out of the box in Section~\ref{sec:framework}. In Section~\ref{sec:models} we describe the process of training and evaluation for new classifiers based on our in-house datasets. We conclude in Section~\ref{sec:conclusisions}.

\section{Bridging TensorFlow and Essentia}
\label{sec:framework}
Our goal is to extend the Essentia framework to support deep learning models with fast inference times and a capability to run on CPUs or acceleration hardware such as GPUs. 
While we could have considered Python-based solutions similar to Madmom~\cite{madmom}, we are interested in an integrated C++ solution 
to take advantage of fast computational speed which is crucial in 
many applications. 
The decision to use TensorFlow instead of other options such as PyTorch~\cite{paszke2017automatic} was motivated by the 
stability of its C API,\footnote{\url{https://www.tensorflow.org/install/lang_c}}
its active development to keep up with the state of the art,
and a huge availability of existing research relying on it.

To this end, we have developed a set of algorithms  
that allow reading frozen models from Protobuf files, generating tensors from 1D or 2D audio representations and running TensorFlow sessions.
The algorithms were implemented with the following design criteria:
\begin{itemize}\compresslist
    \item \textbf{Efficiency.} All dataflow between algorithms for audio feature extraction and model inference should be implemented in C++ without any overhead conversion to Python. We also decided to use TensorFlow frozen models where variables are converted to constants allowing us to remove some training operations.

    \item \textbf{Flexibility.} The deep learning field moves-on fast. Therefore,  generic support for any TensorFlow architecture should be provided. 
    This can be done by loading 
    both the architecture and the weights from external files instead of hard-coding any particular architecture. Importantly, it is also possible to import the models from other frameworks via intermediate formats such as ONNX.

    \item \textbf{Access to intermediate layers.} Sometimes intermediate layers of a model are valuable as they can be used, for example, as features for other tasks. For this reason, it should be possible to extract the output tensors from any layer. 

    \item \textbf{Real-time analysis.} Being able to run computations in real time is one of the key features of Essentia that should be supported by its deep-learning algorithms. 
    The latency and the overall real-time capability ultimately depend on the design of a model, its computational cost for inference, and/or receptive field.   
\end{itemize}

The provided functionality does not include training of the TensorFlow models, only inference. 
Users can be flexible in selecting the way how to train their models as long as they ensure the compatibility of the input features used for training with their implementation in Essentia for inference.
Ideally, users could also use Essentia features on the training stage in order to ensure the best compatibility. 
Many deep learning models proposed in research 
have been trained using features from different software, but they can be also reproduced in Essentia 
as its algorithms are sufficiently configurable for most input audio features. 
For example, in the case of mel-spectrograms, Essentia can reproduce virtually any existing common mel implementation.

Most TensorFlow models can then be 
made compatible with Essentia by freezing and serializing them into Protobuf files. This is a simple process that can be easily done using available Python scripts. 

As an example of the efficiency our framework, we compared inference times for \musicnn{}~\cite{pons2019musicnn} using the original implementation in Python and our algorithms called from Essentia's Python bindings. The original feature extraction time, based on Librosa, took 6.51 seconds compared to 2.30 for Essentia. Loading the model and predicting took 2.07s and 1.66s, respectively. 
In total, considering the extra overhead of dataflow, the difference is 8.60 to 3.34 seconds, meaning that our framework is 2.5 times faster for the entire end-to-end process from loading audio to inference. These time estimations were done averaging 10 trials of analysis of a 3:27 MP3 file on an i7 6700 CPU.

All new algorithms are available as a part of Essentia.
We provide a tutorial with examples of how to install and use the framework, create TensorFlow frozen models and run those models to generate predictions on the example of music auto-tagging.\footnote{\url{https://mtg.github.io/essentia-labs/}}
In addition, we have incorporated a number of state-of-the-art models from audio tagging research into Essentia, listed in Table~\ref{table:soa-models} and made them publicly available on the official website.\footnote{\url{https://essentia.upf.edu/models/}}
We use some of these models in our experiments in Section~\ref{sec:models}.



\begin{table}[t!]
\centering
\footnotesize

\begin{tabular}{llcc}
 \toprule
 Architecture & Dataset & Classes & AUC-PR \\
 \midrule
 \musicnn{} & MSD~\cite{bertin2011million} & 50 & 88.01 \\
 \musicnn{} & MTT~\cite{law2009evaluation} & 50 & 90.69  \\
 VGG & MSD~\cite{bertin2011million} & 50 & 87.67  \\
 VGG & MTT~\cite{law2009evaluation} & 50 & 90.26  \\
 VGG & Audioset~\cite{hershey2016cnn} & 3087 & 91.00 \\
\bottomrule
\end{tabular}
\caption{State-of-the-art CNN models included in Essentia.}
\label{table:soa-models}
\end{table}




\section{Training CNN classifiers for Essentia}
\label{sec:models}

There are many annotated in-house music collections 
that are used extensively in Essentia and a number of related large-scale projects such as AcousticBrainz~\cite{porter_acousticbrainz:_2015}. These collections are summarized in Table~\ref{table:mtgdb-models}. 
Even though their scale is not comparable with many recent datasets, 
they are interesting to work with because they represent a typical use-case of a small amount of data available for a particular application. In addition our intention is to improve the classifiers that have already been used in research and not to challenge the state of the art on any particular task.

In this section we take advantage of these 
datasets to train CNN classifiers in order to improve on the SVM-based models available in Essentia.
Our model creation process is divided in two steps. 
First, we focus on the genre recognition task for which we have additional validation datasets 
to select the best architecture and training strategy. 
Next we use them 
to train classifiers for all our in-house music collections.


\begin{table}
\centering
\footnotesize
    \begin{tabular}{p{20mm}p{40mm}p{13mm}}
    \toprule
    
    Dataset & Classes& Size    \\
    \midrule
    genre-dortmund & alternative, blues, electronic, folkcountry, funksoulrnb, jazz, pop, raphiphop, rock & 1820 exc. \\
    genre-gtzan & blues, classic, country, disco, hip hop, jazz, metal, pop, reggae, rock & 1000 exc. \\
    genre-rosamerica & classic, dance, hip hop, jazz, pop, rhythm and blues, rock, speech  & 400 ft. \\
    \midrule
    mood-acoustic & acoustic, not acoustic & 321 ft. \\
    mood-electronic & electronic, not electronic & 332 ft./exc. \\
    mood-aggressive & aggressive, not aggressive & 280 ft. \\
    mood-relaxed & not relaxed, relaxed & 446 ft./exc. \\
    mood-happy & happy, not happy & 302 exc. \\
    mood-sad & not sad, sad & 230 ft./exc. \\
    mood-party & not party, party & 349 exc. \\
    \midrule
    danceability & danceable, not dancable & 306 ft. \\
    voice/instrumental & voice, instrumental & 1000 exc. \\
    gender & female, male & 3311 ft. \\
    timbre & bright, dark & 3000 exc. \\
    tonal/atonal & atonal, tonal & 345 exc.\\
    \bottomrule
    \end{tabular}
    \caption{In-house music collections (ft.: full tracks, exc.: excerpts).}
    \label{table:mtgdb-models}
\end{table}


\subsection{Architectures, training strategies and experimental setup}
We considered two architectures for building our models
(further details can be found in the original references~\cite{pons2019musicnn,pons2017end,choi2016automatic,hershey2016cnn}):
\begin{itemize}
    \item \textbf{\musicnn{}} is a musically-motivated CNN~\cite{pons2019musicnn}. It uses vertical and horizontal convolutional filters aiming to capture timbral and temporal patterns, respectively. The model contains 6 layers and 787{,}000 parameters.
    \item \textbf{VGG} is an architecture from computer vision based on a deep stack of 3$\times$3 convolutional filters commonly used for audio~\cite{choi2016automatic,hershey2016cnn}. We consider two  different implementations. \textbf{VGG-I} contains 5 layers with 128 filters each. Batch normalization and dropout are applied before each layer~\cite{pons2019musicnn}. The model has 605{,}000 trainable parameters. 
    \textbf{VGG-II} follows the configuration ``E''~\cite{simonyan2014deep} from the original implementation for computer vision, with the difference that the number of output units is set to 3087~\cite{hershey2016cnn}.
    This model has 62 million parameters.
\end{itemize}


\noindent We 
compare transfer learning to the models trained from scratch: 
\begin{itemize}
    \item \textbf{Transfer learning models}. A pre-trained model is loaded and only a small neural network connected to its 
    penultimate layer is trained. The models (\musicnn{}, VGG-I and VGG-II) were previously 
    trained on two audio tagging tasks:
    \begin{itemize}
    	\item \textbf{MSD-train} contains 200{,}000 tracks from the train set of the publicly available Million Song Dataset (MSD)~\cite{bertin2011million} annotated by the 50 Lastfm tags most frequent in the dataset.\footnote{\url{http://millionsongdataset.com/lastfm}}
    	\item \textbf{AudioSet} contains 1.8 million audio clips from Youtube 
    	annotated with the AudioSet taxonomy~\cite{hershey2016cnn}, not specific to music. 
    \end{itemize}
    
    {\musicnn{}} and {VGG-I} are pre-trained on {MSD-train}, while {VGG-II} uses {AudioSet}. We considered two variants of transfer learning back-ends for these models in a preliminary experiment:
    \textit{(A)} one fully connected output layer of $n$ units and \textit{(B)} two fully connected layers of 100 and $n$ units, respectively, where $n$ is the number of classes in the employed dataset. 
The variant A provided the best results for \musicnn{} and VGG-I, while the variant B gave the best results for VGG-II. We used these best configurations for each model in the rest of our study.


    

    \item \textbf{Models trained from scratch}. The parameters of 
    \musicnn{} and VGG-I 
    are randomly initialized and all the layers are trained. 

\end{itemize}


All our CNNs were trained on mel-spectrograms. For the models trained from scratch we used the implementation in Essentia with 96 bands.
In the case of transfer learning, we used 96 bands for \musicnn{} and VGG-I, and 64 bands for VGG-II. We opted for the feature extractors used by the authors of the pre-trained models, but we re-implemented those mel-spectrograms for inference.



To estimate the accuracy of each model we conduct a stratified 5-fold cross-validation, where each training split is further divided into 80\% train and 20\% validation subsets. After this, to take advantage of all data possible, the final CNN models that 
we evaluate on external datasets are trained using the 80\% of the entire data (20\% is kept for validation). 

The models are trained on mini-batches of 32 samples. Each sample is a randomly selected segment of 3 seconds from a different track of the training set.
SGD employing Adam is used as the optimizer.
The number of epochs is 600 for the models trained from scratch. The transfer learning models are trained for 150 epochs, as those models require less iterations to converge.
All the models are initialized with a learning rate of 0.001.
If the loss obtained on the validation set has not decreased for the last 75 epochs, the learning rate is reduced by half.

The baseline for our experiments is comprised of the SVM classifiers available in Essentia.\footnote{We used the latest Essentia 2.1-beta5 version.} They rely on a combination of low-, mid- and high-level music audio features describing timbre, rhythm and tonality~\cite{porter_acousticbrainz:_2015}. The best parameters for the SVMs are found in a grid search in the 5-fold cross-validation, and the final SVM models that we evaluate are trained on the entire data.\footnote{\url{https://essentia.upf.edu/documentation/FAQ.html}} 

We used standard TensorFlow routines in Python for training and then stored the models into Protobuf files to be used in Essentia.




\begin{table*}[t]
    \centering
    \footnotesize

    \begin{tabular}{ccc cc c c}
        \toprule
       Genre dataset & Baseline      & \multicolumn{2}{c}{Trained from scratch} & \multicolumn{3}{c}{Transfer learning} \\ 
       \cmidrule(lr){2-2} \cmidrule(lr){3-4} \cmidrule(lr){5-7}
        & SVM & \musicnn{} & VGG I & \musicnn{} (MSD-train) & VGG-I (MSD-train) & VGG-II (Audioset) \\
        \midrule

        \multicolumn{7}{c}{5-fold cross-validation} \\ \midrule
        dortmund &  0.42$\pm$.01 & 0.40$\pm$.03 & 0.43$\pm$.02 & 0.51$\pm$.02 & 0.47$\pm$.02 & \textbf{0.52}$\pm$\textbf{.02} \\ 
        gtzan  &  0.74$\pm$.01 & 0.83$\pm$.02 & 0.82$\pm$.01 & 0.81$\pm$.03 & 0.83$\pm$.01 & \textbf{0.86}$\pm$\textbf{.02} \\
        rosamerica  &  0.86$\pm$.02 & 0.93$\pm$.03 & 0.88$\pm$.02 & 0.92$\pm$.02 & 0.92$\pm$.02 & \textbf{0.94}$\pm$\textbf{.02} \\
        \midrule
        \multicolumn{7}{c}{Cross-collection evaluation on MSD-test} \\ \midrule
        dortmund & 0.36 & 0.35 & 0.35 & 0.39 & 0.37 & \textbf{0.42} \\ 
        gtzan & 0.28 & 0.34 & 0.49 & 0.51 & 0.50 & \textbf{0.54} \\
        rosamerica & 0.44 & 0.44 & 0.46 & 0.52 & 0.50 & \textbf{0.54} \\

        \midrule
        \multicolumn{7}{c}{Cross-collection evaluation on MTG-Jamendo-test} \\ \midrule
        dortmund &0.18 & 0.32 & 0.35 & 0.37 & 0.35 & \textbf{0.41} \\
        gtzan & 0.11 & 0.35 & 0.37 & 0.40 & 0.37 & \textbf{0.44} \\
        rosamerica & 0.37 & 0.43 & 0.44 & 0.46 & 0.44 & \textbf{0.48} \\
        \bottomrule
    \end{tabular}
    
    \caption{Cross-collection evaluation results. The best balanced accuracies are marked in bold.}
    \label{table:cce}
\end{table*}

\subsection{Evaluation on genre recognition tasks}


Given the small size of our datasets, overfitting can be an issue and the results of the 5-fold cross-validation can be unreliable. 
For this reason, we conduct a cross-collection evaluation that consists in evaluating the models on an independent source of music and annotations 
following the methodology proposed in~\cite{bogdanov2016cross}. 
This allows us to identify the model architecture and training strategy with the best generalization capabilities.

Unfortunately, we are lacking such external datasets 
to evaluate all our classifiers, but we are able to do it for the task of genre classification for which we have three datasets: \textit{genre-dortmund}, \textit{genre-gtzan} and \textit{genre-rosamerica}.
As external data sources we use two datasets, 
both containing tag annotations including genres:
\begin{itemize}\compresslist
    \item \textbf{MSD-test} is the test set of 28{,}000 tracks from the MSD dataset with Lastfm tags. Note that MSD has been also used for 
    the pre-trained \musicnn{} and VGG-I models, 
    but 
    they were trained on the train split and there is no overlap.

    \item \textbf{MTG-Jamendo-test} is the \textit{split-0} test set of 11{,}000 tracks from the MTG-Jamendo dataset for music tagging~\cite{bogdanov2019jamendo}.
    \footnote{\url{https://mtg.github.io/mtg-jamendo-dataset}} 
\end{itemize}

Following~\cite{bogdanov2016cross}, we took advantage of the taxonomy used by the Lastgenre plugin for Beets\footnote{\url{http://beets.io}} 
to generate ground-truth genre labels from the tags in MSD-test and MTG-Jamendo-test. We only considered tracks with one or more tags matching an element in the taxonomy. Those tags were mapped to its 
parent in the hierarchy (e.g., ``progressive rock'' to ``rock''), unless there was a direct match to one of the classes of our classifiers. 
The resulting genre annotations are multi-label, and to evaluate each group of classifiers (corresponding to one of our in-house datasets) we use the subset of tracks that have a ground-truth label matching one of the classes. That is, we only give them music by genres they can theoretically predict. 
A prediction is considered correct if it matches one of the labels of the track.




Table~\ref{table:cce} contains the balanced accuracies obtained by each architecture and training strategy 
in both the 5-fold cross-validation and cross-collection evaluation on MSD-test and MTG-Jamendo-test. 
These accuracies are computed by averaging the individual recall values obtained for each class. For the cross-validation results we indicate the standard deviation of the balanced accuracies across folds. 
Our results show that transfer learning models, in particular VGG-II with AudioSet, consistently outperform the SVMs and the CNNs trained from scratch. 
Interestingly, the AudioSet model is not specifically trained for music content, but it is still capable to get the best results, potentially due to its training data size and complexity.




\subsection{Training Essentia models}

As genre classification is a complex problem, we can expect that the conclusions of the previous section 
will also benefit the rest of our classification tasks.
Therefore, we use the winning architecture and training strategy from the previous experiment to generate new models for the rest of the tasks including mood classification and other high-level music description. 

Again we used our SVM classifiers as a baseline. 
For further quality assessment of our models, 
we decided to manually annotate a subset of 
MTG-Jamendo-test  
by the classes of our classifiers. 
This subset contains approximately 1{,}000 tracks selected by a stratified approach~\cite{sechidis2011stratification, 2017arXiv170201460S} in order to maximize the variety of music according to the associated tags.
The annotations were performed solely utilizing the labels from 
the taxonomies of our in-house datasets.
The final number of tracks used to evaluate each model varies (from 599 to 1000) as we discarded the tracks that could not be matched to any taxonomy class by the annotators.
We use these annotations as a ground truth to compare predictions by the transfer-learning models to the SVMs in terms of balanced accuracies.

Table \ref{table:inhouseacc} presents the results for all tasks, including 5-fold cross-validation on the original datasets used for training as well as evaluation on our manually annotated subset of MTG-Jamendo-test.
As we can see, 
VGG-II with AudioSet 
leads to improvement in the mean accuracies over SVM baseline in the 5-fold cross-validation, however the difference is not statistically significant in many cases. Meanwhile, the results on the manually annotated subset of MTG-Jamendo-test show that our CNN models perform better except for the models for mood-party, danceability, gender and timbre.

It is important to note that
we did not take much care on the optimization of the hyper-parameters of the models, still getting decent improvements on a number of the datasets and opening possibilities for future work.  
Overall, we 
can see 
better generalization 
of the CNN models 
in the cross-collection evaluation for many of the datasets.

\begin{table}[h!]
\centering
\footnotesize
    \begin{tabular}{lcccc}
    \toprule
          &\multicolumn{2}{c}{5-fold cross-validation} & \multicolumn{2}{c}{MTG-Jamendo-test} \\ 
       \cmidrule(lr){2-3} \cmidrule(lr){4-5}

    Model & SVM & VGG-II & SVM & VGG-II  \\ 
          &     &   (AudioSet) &   & (AudioSet) \\
    \midrule
    genre-dortmund  & 0.42$\pm$.01 & \textbf{0.52}$\pm$.02 & 0.19 & 0.48 \\  
    genre-gtzan & 0.77$\pm$.03 & \textbf{0.86}$\pm$.02 & 0.14 & 0.58 \\ 
    genre-rosamerica & 0.86$\pm$.02 & \textbf{0.94}$\pm$.02 & 0.47 & 0.53 \\ 
\midrule


    mood-acoustic   & 0.93$\pm$.02 & 0.94$\pm$.03 & 0.75 & 0.82 \\ 
    mood-electronic & 0.83$\pm$.03 & \textbf{0.93}$\pm$.03 & 0.70 & 0.83 \\ 
    mood-aggressive & 0.97$\pm$.02 & 0.98$\pm$.02 & 0.67 & 0.74 \\ 
    mood-relaxed    & 0.89$\pm$.04 & 0.89$\pm$.03 & 0.60 & 0.73 \\ 

    mood-happy & 0.81$\pm$.04 & 0.86$\pm$.04 & 0.60 & 0.71 \\ 
    mood-sad & 0.88$\pm$.06 & 0.89$\pm$.02 & 0.65 & 0.72 \\ 
    mood-party & 0.88$\pm$.05 & 0.91$\pm$.06 & 0.77 & 0.76 \\ 
    \midrule

    danceability & 0.90$\pm$.03 & \textbf{0.94}$\pm$.02 & 0.77 & 0.72 \\ 
    voice/instrumental & 0.93$\pm$.01 & \textbf{0.98}$\pm$.01 & 0.72 & 0.87 \\ 
    gender & 0.88$\pm$.01 & 0.84$\pm$.01 & 0.44 & 0.40 \\ 
    timbre & 0.94$\pm$.06 & 0.93$\pm$.01 & 0.54 & 0.52\\ 
    tonal/atonal & 0.98$\pm$.01 & 0.97$\pm$.03 & 0.60 & 0.66 \\ 
    \bottomrule
    \end{tabular}
    \caption{Balanced accuracies for 5-fold cross-validation and evaluation on a manually annotated subset of MTG-Jamendo-test.  Statistically significant improvements over the SVMs according to an independent samples t-test ($P > 0.05$) are marked in bold.}
    \label{table:inhouseacc}
\end{table}



\section{Conclusions}\label{sec:conclusisions}
We have presented our development effort to add support for generic TensorFlow 
models in Essentia, a C++ library for audio and music analysis with Python bindings,
being the first effort of its kind to integrate arbitrary deep learning models into an MIR library.
The new functionality for using such models is designed to be fast, easy and flexible, and it is especially attractive for applications requiring computational efficiency, such as large-scale analysis on millions of tracks, real-time processing, or inference on weak devices. 

We provide a number of CNN audio tagging models, 
ported from Python implementations made by other researchers and our own classifier models trained using in-house datasets.
For the latter models we apply transfer learning techniques that outperform previous Essentia classifiers based on SVMs.
All of these models are publicly available for researchers and practitioners, and we plan to add more models in the future.

\newpage

\bibliographystyle{IEEEbib}
\bibliography{text}

\end{document}